\documentclass[aps,pra]{revtex4}
\usepackage{graphicx}

\begin{document}

\title{Less eavesdropping losses induce more eavesdropping information gain
\thanks{E-mail: Zhangzj@wipm.ac.cn}}

\author{
Zhanjun Zhang$^1$, Zhongxiao Man$^1$ and Yong Li$^2$  \\
{\footnotesize $^1$ Wuhan Institute of Physics and Mathematics,
TheChinese Academy of Sciences, Wuhan 430071, China} \\
{\footnotesize $^2$ Department of Physics, Center China Normal
University, Wuhan 430079, China} \\
{\footnotesize E-mail: Zhangzj@wipm.ac.cn}}
\date{\today}

\maketitle

\begin{minipage}{400pt}
The eavesdropping scheme proposed by W\'{o}jcik [Phys. Rev. Lett.
{\bf 90},157901(2003)] on the quantum communication protocol of
Bostr\"{o}m and Felbinger [Phys. Rev. Lett. {\bf 89},
187902(2002)] is improved by constituting a new set of attack
operations. The improved scheme only induces half of the
eavesdropping losses in W\'{o}jcik's scheme, therefore, in a
larger domain of the quantum channel transmission efficiency
$\eta$, i.e., [0,75\%], the eavesdropper Eve can attack all the
transmitted bits. Comparing to W\'{o}jcik's scheme, in the
improved scheme the eavesdropping (legitimate) information gain
does not vary in the $\eta$ domain of [0, 50\%], while in the
$\eta$ domain of (50\%, 75\%] the less eavesdropping losses induce
more eavesdropping information gains, for Eve can attack {\it all}
the transmitted bits and accordingly eavesdropping information
gains do {\it not} decrease. Moreover, for the
Bostr\"{o}m-Felbinger protocol, the insecurity upper bound of
$\eta$ presented by W\'{o}jcik is pushed up in the this paper,
that is, according to W\'{o}jcik's eavesdropping scheme, the
Bostr\"{o}m-Felbinger protocol is not secure for transmission
efficiencies lower than almost $60\%$, while according to the
improved scheme, it is not secure for transmission efficiencies
lower than almost $80\%$.  \\

PACS number(s): 03.67.Hk, 03.65.Ud \\
\end{minipage} \\

Quantum key distribution (QKD) is an ingenious application of
quantum mechanics, in which two remote legitimate users (Alice and
Bob) establish a shared secret key through the transmission of
quantum signals. Much attention has been focused on QKD after the
pioneering work of Bennett and Brassard published in 1984 [1].
Till now there have been many theoretical QKDs [2-20]. Different
from the QKDs, the deterministic secure direct communication
protocol is to transmit directly the secret messages without first
generating QKD to encrypt them. Hence it is very useful and
usually desired, especially in some urgent time. However, the
deterministic secure direct communication is more demanding on the
security than QKDs. Therefore, only recently a few of
deterministic secure direct protocols have been proposed [21-24].
One of them is the famous Bostr\"{o}m-Felbinger protocol [22],
which allows the generation of a deterministic key or even direct
secret communication. In Ref [22] the protocol has been claimed to
be secure and experimentally feasible. However, since the security
of the Bostr\"{o}m-Felbinger protocol can be impaired as far as
considerable quantum losses are taken into account, very recently
W\'{o}jcik has presented an undetectable eavesdropping scheme on
the Bostr\"{o}m-Felbinger protocol[25]. W\'{o}jcik's eavesdropping
scheme induces the eavesdropping losses at the level of 50\% and
the anticorrelation of the state of the home photon (kept by Bob,
the legitimate receiver of secret messages) with that of the
travel photon (sent by Bob to Alice, the sender of secret
messages). If the transmission efficiency $\eta$ of the quantum
channel is not taken into account, the probability of the
eavesdropper (i.e., Eve) being detected is zero due to the
anticorrelation. However, in the case of the considerable quantum
channel losses, it is possible for legitimate users to detect the
eavesdropping by observing the quantum channel losses. That is,
although Eve can attack all the transmitted bits and the
eavesdropping losses can be hidden in the channel losses when
$\eta \leq 50\%$, if she attacks all the transmitted bits when
$\eta> 50\%$, then the eavesdropping losses is greater than the
channel losses and accordingly the legitimate users can find Eve
in the line by observing the channel losses. In fact, when $\eta >
50\%$, it is still possible for Eve to avoid the legitimate users'
detection, for she can eavesdrop only the fraction $\mu=2(1-\eta)$
of the transmitted bits to induce less eavesdropping losses, which
can be completely hidden in the quantum channel losses. Hence, in
[25] it is concluded that the Bostr\"{o}m-Felbinger protocol is
not secure for transmission efficiencies lower than almost $60\%$,
i.e., the insecurity upper bound of $\eta$ is 60\%. Nonetheless,
in the case of $\eta > 50\%$, the eavesdropping (legitimate)
information gain will surely decrease (increase) due to
eavesdropping only a fraction of the whole transmitted bits. Then
it is intriguing to ask, in a low-loss quantum channel (i.e., a
high transmission efficiency channel), whether Eve can have more
information gain in a way of reducing the eavesdropping losses
such that she is able to eavesdrop all the transmitted bits in a
larger domain of $\eta$. If so, then the insecurity upper bound of
$\eta$ presented by W\'{o}jcik can be pushed up. To address the
question, in this brief report, we improve the the W\'{o}jcik's
eavesdropping scheme [25] by constituting a new set of attack
operations. Our improved eavesdropping scheme indeed induces less
eavesdropping losses than that in [25], therefore, in a larger
domain of $\eta$, Eve can attack all the transmitted bits. Since
when $\eta=1$ the eavesdropping (legitimate) information gain does
not decrease (increase) with the decrease of the eavesdropping
losses, the larger domain in which Eve can attack all the
transmitted bits means that less eavesdropping losses may induce
more eavesdropping information gain. Hence, for the
Bostr\"{o}m-Felbinger protocol, the insecurity upper bound of the
transmission efficiency presented by W\'{o}jcik can be pushed up.
One will see these later.

Let us start with the brief description of the
Bostr\"{o}m-Felbinger protocol [22]. Bob prepares two photons in
the entangled state $|\Psi^+ \rangle = (|0\rangle |1\rangle +
|1\rangle |0\rangle ) / \sqrt{2}$ of the polarization degrees of
freedom. He stores one photon (home photon) in his lab and sends
Alice the other one (travel photon) via a quantum channel. After
receiving the travel photon Alice randomly switches between the
control mode and the message mode. In the control mode Alice
measures the polarization of the travel photon first and then
announces publicly the measurement result and the measurement
basis she used. After knowing Alice's announcement Bob also
switches to the control mode to measure the home photon in the
same basis as that Alice used. Then he compares both measurement
results. They should be perfectly anticorrelated in the absence of
Eve. Therefore, the appearance of identical results is considered
to be the evidence of eavesdropping, and if it occurs the
transmission is aborted. In the other case, the transmission
continues. In the message mode, Alice performs the $Z_t^j  (j \in
\{0,1\})$ operation on the travel photon to encode $j$ and sends
it back to Bob, where $Z=|0\rangle \langle 0| - |1\rangle \langle
1|$. After receiving the travel photon Bob measures the state of
both photons in the Bell basis to decode the $j=0(1)$
corresponding to the $|\Psi^+\rangle(|\Psi^-\rangle)$ result.

Now along the line of W\'{o}jcik's eavesdropping scheme [25], we
present detailedly the improved scheme, which never produces the
identical results of the measurements performed by Bob and Alice
in the control mode also but induces less eavesdropping losses
than that in [25]. Let us first consider an ideal quantum channel
(i.e., $\eta=1$). Obviously, Eve has no access to the home photon
but can manipulate the travel photon while it goes from Bob to
Alice and back from Alice to Bob. Eve uses two auxiliary spatial
modes $x,y$. She prepares a photon in the state $|0\rangle$ and
lets the other one be an empty mode, e.g., in the state $|{\rm
vac} \rangle_x |0\rangle_y$. Accordingly, the state of the whole
system including the entangled photon pair is
\begin{eqnarray}
|initial \rangle = |\Psi^+ \rangle_{ht}|{\rm vac} \rangle_x
|0\rangle_y.
\end{eqnarray}
When Bob sends the travel photon to Alice, Eve attacks the quantum
channel by manipulating the travel photon through an unitary
operation (referred as to be the $B-A$ attack hereafter) as
follow,
\begin{eqnarray}
T=N_{xy}C_{ytx}N_{ty}C_{txy} H_y,
\end{eqnarray}
where $N$ represents the CNOT gate [25], $H$ is the Hadamard gate,
and $C$ stands for the so-called three-mode CPBS gate [25], which
is constructed by CNOT gates and a polarizing beam splitter
transmitting (reflecting) photons in the state $|0\rangle (
|1\rangle)$. When acting on the initial state, the $B-A$ attack
transforms the whole system to the state $|B-A\rangle=T|initial
\rangle$ of the form
\begin{eqnarray}
|B-A\rangle= \frac{1}{2}|0\rangle_h(|{\rm
vac}\rangle_t|1\rangle_x|0\rangle_y + |1\rangle_t|1\rangle_x|{\rm
vac}\rangle_y)
\nonumber \\
+\frac{1}{2}|1\rangle_h(||0\rangle_t{\rm vac}\rangle_x |1\rangle_y
+ |0\rangle_t|0\rangle_x|{\rm vac}\rangle_y).
\end{eqnarray}
Suppose that Alice now switches to the control mode and measures
the state of the mode $t$.  According to equation 3, one can see
that after the B-A attack Alice will detect no photon with the
probability $1/4$ or with the probability of $3/4$ a photon whose
state is perfectly anticorrelated with the state of the home
photon. Therefore, the probability of eavesdropping detection
based on the correlation observation equals zero. This point is
completely same as that in Ref.[25]. Moreover, from equation 3 one
can see that the $B-A$ attack in the present eavesdropping scheme
also induces the eavesdropping losses. However, it is worthy to be
mentioned that the eavesdropping losses level (25\%) in the
present scheme is only half of that (50\%) in Ref.[25]. This
implies that, comparing to the W\'{o}jcik's eavesdropping scheme,
in the present eavesdropping scheme, the domain in which Eve can
attack all the transmitted bits is enlarged to be [0, 75\%] from
the [0, 50\%] in Ref.[25]. Nonetheless, this does not certainly
mean that, the insecurity upper bound of transmission efficiency
presented by W\'{o}jcik can be pushed up, for now we still do not
know the variation of the eavesdropping (legitimate) information
gain. Let us now analyze the performance of the scheme in the case
of Alice operating in the message mode. After Alice performs the
$Z^j$ operation and sends the travel photon back to Bob, Eve
performs her second attack (named as the $A-B$ attack hereafter)
on the travel photon. The $A-B$ attack consists of the unitary
operation $T^{-1}$. After the $A-B$ attack, the corresponding
state of the whole system is
\begin{eqnarray}
|A-B\rangle &=& T^{-1}Z_t^j|B-A\rangle \nonumber \\
&=&\frac{1}{\sqrt{2}}(|0\rangle_h|1\rangle_t|j\rangle_y +
|1\rangle_h|0\rangle_t|0\rangle_y)|{\rm vac} \rangle_x \nonumber
\\ &=&\frac{1}{2}(|\Psi^+\rangle_{ht}|j\rangle_y
+|\Psi^-\rangle_{ht}|j\rangle_y +|\Psi^+\rangle_{ht}|0\rangle_y
-|\Psi^-\rangle_{ht}|0\rangle_y )|{\rm vac}\rangle_x.
\end{eqnarray}
The final step of the eavesdropping scheme is a measurement of
polarization performed on the $y$ photon, the result is denoted by
$k$. The result of Bob's Bell-state measurement on both photons is
denoted by $m=0 (1)$ corresponding to the
$|\Psi^+\rangle_{ht}(|\Psi^-\rangle_{ht})$ state. Let $P(k,m|j)$
be the conditional probability of possible measurement outputs of
$|A-B\rangle$ for a given value of $j$, then the only nonzero
probabilities are,
\begin{eqnarray}
P_I(0,0|0)=1, \,\,\,\,\,
P_I(0,0|1)=P_I(0,1|1)=P_I(1,0|1)=P_I(1,1|1)=1/4.
\end{eqnarray}
Assume that in Alice's secret messages, the occupation
possibilities of the '0' and the '1' bits are $c_0$ and $1-c_0$
respectively, then one can work out the mutual information between
any two parties,
\begin{eqnarray}
I_{AE}=I_{AB}=c_0-\frac{1}{2}[(1-c_0)\log_2(1-c_0)+(1+c_0)\log_2(1+c_0)],\\
I_{BE}=-(1+c_0)\log_2(1+c_0)+\frac{1-c_0}{4}\log_2(1-c_0)+\frac{1+3c_0}{4}\log_2(1+3c_0).
\end{eqnarray}
Specifically, when $c_0=1/2$, then
\begin{eqnarray}
I_{AE}=I_{AB}=\frac{3}{4}\log_2\frac{4}{3} \approx0.311 \nonumber \\
I_{BE}=1-\frac{3}{2}\log_23+\frac{5}{8}\log_25\approx 0.074.
\end{eqnarray}

Same as W\'{o}jcik's eavesdropping scheme, the present scheme is
also not symmetric. If after the $A-B$ attack Eve performs an
additional unitary operation $S=X_tZ_tN_{ty}X_t$ with the
probability of $1/2$, where $X$ is an negation, then the asymmetry
can be removed. If the $S$ is performed, the final state of the
whole system is transformed into
\begin{eqnarray}
|A-B\rangle^{(s)} &=& ST^{-1}Z_t^j|B-A\rangle \nonumber \\
&=&\frac{1}{\sqrt{2}}(|0\rangle_h|1\rangle_t|j\rangle_y -
|1\rangle_h|0\rangle_t|1\rangle_y)|{\rm vac} \rangle_x \nonumber
\\ &=&\frac{1}{2}(|\Psi^+\rangle_{ht}|j\rangle_y
+|\Psi^-\rangle_{ht}|j\rangle_y -|\Psi^+\rangle_{ht}|1\rangle_y
+|\Psi^-\rangle_{ht}|1\rangle_y )|{\rm vac}\rangle_x.
\end{eqnarray}
Then the only nonzero $P(k,m|j)$'s are,
\begin{eqnarray}
P^{(s)}_I(0,0|0)=P^{(s)}_I(0,1|0)=P^{(s)}_I(1,0|0)=P^{(s)}_I(1,1|0)=1/4,
\,\,\,\,\,P^{(s)}_I(1,1|1)=1.
\end{eqnarray}
Considering the assumption that in Alice's secret messages the
occupation possibilities of the '0' and the '1' bits are $c_0$ and
$1-c_0$ respectively, one can work out the mutual information
between any two parties,
\begin{eqnarray}
I_{AE}^{(s)}=I_{AB}^{(s)}=1-c_0-\frac{1}{2}[c_0\log_2c_0+(2-c_0)\log_2(2-c_0)],\\
I_{BE}^{(s)}=-(2-c_0)\log_2(2-c_0)+\frac{c_0}{4}\log_2c_0+\frac{4-3c_0}{4}\log_2(4-3c_0).
\end{eqnarray}
Specifically, when $c_0=1/2$, then
\begin{eqnarray}
I_{AE}^{(s)}=I_{AB}^{(s)}=\frac{3}{4}\log_2\frac{4}{3} \approx0.311 \nonumber \\
I_{BE}^{(s)}=1-\frac{3}{2}\log_23+\frac{5}{8}\log_25\approx 0.074.
\end{eqnarray}
Incidentally, it is easily found that, when $c_0 \neq 1/2$ (Note
that $c_0=0$ or $c_0=1$ is meaningless), $I_{AE} \neq
I_{AE}^{(s)}$, $I_{AB} \neq I_{AB}^{(s)}$ and $I_{BE} \neq
I_{BE}^{(s)}$. Same as that in Ref.[25], later we only consider
the case of $c_0=1/2$.

According to the viewpoint in Ref.[25], that is, since Eve knows
exactly when each of the $S$ operations has been performed, the
symmetrization procedure does not reduce the mutual information
between Alice and Eve while it disturbs the communication between
Alice and Eve in such a way the mutual information between Alice
and Bob is reduced, in the present scheme after the $S$ operations
the mutual information between Alice and Bob is also reduced to be
$ I_{AB}=\frac{3}{4}\log_23-1 \approx0.189$.

By the way, one can easily find that in the present eavesdropping
scheme the quantum bit error rate induced by the eavesdropping is
also at the same level of $1/4$ as that in Ref.[25].

Thus far, we have improved W\'{o}jcik's eavesdropping scheme. One
can see that the improved scheme is almost same as W\'{o}jcik's
eavesdropping scheme except for the induced eavesdropping losses.
Hence, all the discussions in Ref.[25] except for those related to
the eavesdropping losses are also suitable for the present paper
and we will not repeat them. Now let us discuss a very important
property related with the eavesdropping losses. According to our
above calculations of the mutual information, one can see that,
when $\eta=1$, comparing to W\'{o}jcik's eavesdropping scheme, the
eavesdropping (legitimate) information gain does not decrease
(increase) with the decrease of the eavesdropping losses in the
improved scheme. As mentioned before, in the improved scheme the
$\eta$ domain in which Eve can attack all the transmitted bits is
enlarged to be [0, 75\%] from the [0, 50\%] in Ref.[25]. In the
$\eta$ domain of (50\%, 75\%], the eavesdropping information gain
does not decrease in the improved scheme but in W\'{o}jcik's
eavesdropping scheme. Thus, in this sense, one can say that the
less eavesdropping losses do induce more eavesdropping information
gain. Moreover, in the improved scheme, when $\eta$ increases from
75\%, though the mutual information between Alice and Eve (Bob)
will decrease (increase) for Eve can only attack a fraction of the
whole transmitted bits, the $I_{AE}$ is still able to exceed the
$I_{AB}$ up to almost 80\% transmission efficiency (cf. Fig.1).

In summary, we have improved W\'{o}jcik's eavesdropping scheme on
the Bostr\"{o}m-Felbinger quantum communication protocol. The
improved scheme only induces half of the eavesdropping losses in
W\'{o}jcik's scheme and accordingly in the $\eta$ domain of [0,
75\%] Eve can attack all the transmitted bits. When $\eta=1$ the
eavesdropping (legitimate) information gain does not accordingly
decrease (increase). Hence, in the $\eta$ domain of (50\%,75\%],
the less eavesdropping losses do induce more eavesdropping
information gain. Moreover, as for as the Bostr\"{o}m-Felbinger
protocol is concerned, the insecurity upper bound of the
transmission efficiency has been pushed up from the 60\% in
W\'{o}jcik's scheme to 80\% in the present scheme.

Zhang thanks to Prof. Baiwen Li for his encouragement. This work
is funded by the National Science Foundation of China under No.10304022.\\

\vskip 2cm
\begin{figure}
\begin{center}
\includegraphics[width=1.0\textwidth]{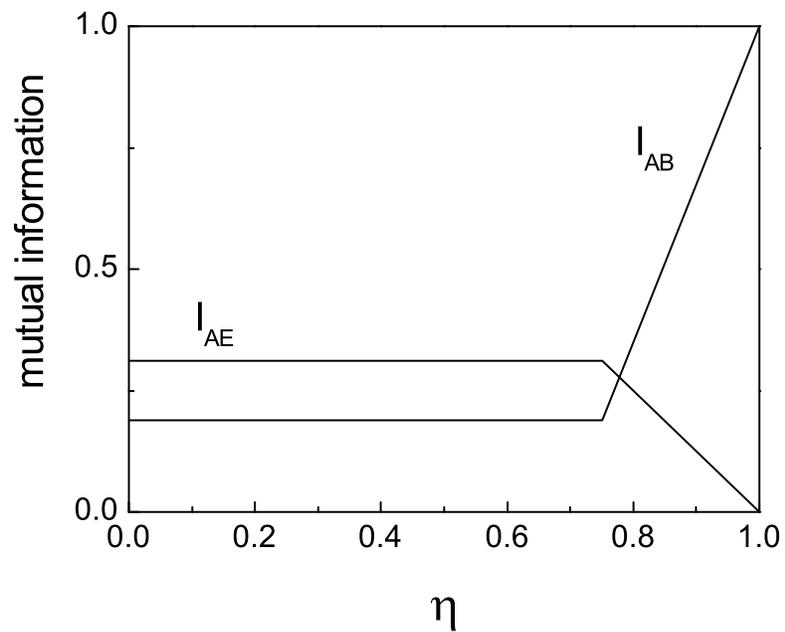}
\vskip -2cm \caption{Mutual information between Alice and Eve
(Bob) $I_{AE}$($I_{AB}$) as a function of quantum channel
transmission efficiency $\eta$. According to the figure 4 in
Ref.[25], when $\eta$ increases form 75\%,  $I_{AB}$ in this
figure should increase not linearly but more slowly. We simplify
it just because we only want to estimate an approximate upper
bound.} \label{fig1}
 \end{center}
\end{figure}

\newpage

\noindent [1] C. H. Bennett and G. Brassard, in {\it Proceedings
of the IEEE International Conference on Computers, Systems and
Signal Processings, Bangalore, India} (IEEE, New York, 1984),
p175.

\noindent[2] N. Gisin, G. Ribordy, W. Tittel, and H. Zbinden, Rev.
Mod. Phys. {\bf 74},145 (2002).

\noindent[3] A. K. Ekert, Phys. Rev. Lett. {\bf67}, 661 (1991).

\noindent[4] C. H. Bennett, Phys. Rev. Lett. {\bf68}, 3121
 (1992).

\noindent[5] C. H. Bennett, G. Brassard, and N.D. Mermin, Phys.
Rev. Lett. {\bf68}, 557(1992).

\noindent[6] L. Goldenberg and L. Vaidman, Phys. Rev. Lett.
{\bf75}, 1239  (1995).

\noindent[7] B. Huttner, N. Imoto, N. Gisin, and T. Mor, Phys.
Rev. A {\bf51}, 1863 (1995).

\noindent[8] M. Koashi and N. Imoto, Phys. Rev. Lett. {\bf79},
2383 (1997).

\noindent[9] W. Y. Hwang, I. G. Koh, and Y. D. Han, Phys. Lett. A
{\bf244}, 489 (1998).

\noindent[10] P. Xue, C. F. Li, and G. C. Guo,  Phys. Rev. A
{\bf65}, 022317 (2002).

\noindent[11] S. J. D. Phoenix, S. M. Barnett, P. D. Townsend, and
K. J. Blow, J. Mod. Opt. {\bf42}, 1155 (1995).

\noindent[12] H. Bechmann-Pasquinucci and N. Gisin, Phys. Rev. A
{\bf59}, 4238 (1999).

\noindent[13] A. Cabello, Phys. Rev. A {\bf61},052312 (2000);
{\bf64}, 024301 (2001).

\noindent[14] A. Cabello, Phys. Rev. Lett. {\bf85}, 5635 (2000).

\noindent[15] G. P. Guo, C. F. Li, B. S. Shi, J. Li, and G. C.
Guo, Phys. Rev. A {\bf64}, 042301 (2001).

\noindent[16] G. L. Long and X. S. Liu, Phys. Rev. A {\bf65},
032302 (2002).

\noindent[17] F. G. Deng and G. L. Long, Phys. Rev. A {\bf68},
042315 (2003).

\noindent[18]J. W. Lee, E. K. Lee, Y. W. Chung, H. W. Lee, and J.
Kim, Phys. Rev. A {\bf 68}, 012324 (2003).

\noindent[19] Daegene Song, Phys. Rev. A {\bf69}, 034301(2004).

\noindent[20] X. B. Wang, Phys. Rev. Lett. {\bf 92}, 077902
(2004).

\noindent[21] A. Beige, B. G. Englert, C. Kurtsiefer, and
H.Weinfurter, Acta Phys. Pol. A {\bf101}, 357 (2002).

\noindent[22] Kim Bostrom and Timo Felbinger, Phys. Rev. Lett.
{\bf89}, 187902 (2002).

\noindent[23] F. G. Deng, G. L. Long, and X. S. Liu, Phys. Rev. A
{\bf68}, 042317 (2003).

\noindent[24]F. G. Deng and G. L. Long,  Phys. Rev. A {\bf x},
(2004).

\noindent [25] A. Wojcik, Phys. Rev. Lett. {\bf90}, 157901 (2003).

\end{document}